\documentclass[sigconf]{acmart}
\renewcommand\footnotetextcopyrightpermission[1]{}

\usepackage{algorithm}
\usepackage{algorithmic}
\usepackage{enumitem}
\usepackage{subcaption}
\usepackage{multirow}
\newcounter{requirement-c}
\usepackage{tcolorbox}
\newtcolorbox{textsample}[1][]{
  colback=white, 
  colframe=black, 
  coltitle=black, 
  sharp corners, 
  boxrule=0.7pt, 
  left=3pt, 
  right=3pt,
  top=3pt,
  bottom=3pt,
  before skip=4pt, 
  after skip=4pt, 
  toptitle=2pt, 
  bottomtitle=2pt, 
  fonttitle=\normalsize\bfseries,
}
\usepackage{cleveref}
\crefformat{section}{#2Section~#1#3}
\crefformat{figure}{#2Figure~#1#3}
\crefformat{algorithm}{#2Algorithm~#1#3}
\crefformat{table}{#2Table~#1#3}
\crefformat{equation}{#2Equation~#1#3}
\crefformat{definition}{#2Definition~#1#3}
\crefformat{appendix}{#2Appendix~#1#3}

\usepackage{makecell} 

\newcommand{\dx}{d_\mathcal{X}}
\newcommand{\noise}{\boldsymbol{\eta}}
\newcommand{\norm}[1]{\lVert{#1}\rVert}
\newcommand{\embmod}{\phi}
\newcommand{\RIGHTCOMMENT}[1]{\hfill \(\triangleright\) #1}

\begin{document}

\title{Preempting Text Sanitization Utility in Resource-Constrained Privacy-Preserving LLM Interactions}

\author{Robin Carpentier}
\email{robin.carpentier@mq.edu.au}
\orcid{0000-0003-1369-2248}
\affiliation{%
	\institution{Macquarie University}
	\city{Sydney}
	\country{Australia}
}
\author{Benjamin Zi Hao Zhao}
\email{ben\_zi.zhao@mq.edu.au}
\orcid{0000-0002-2774-2675}
\affiliation{%
	\institution{Macquarie University}
	\city{Sydney}
	\country{Australia}
}
\author{Hassan Jameel Asghar}
\email{hassan.asghar@mq.edu.au}
\orcid{0000-0001-6168-6497}
\affiliation{%
	\institution{Macquarie University}
	\city{Sydney}
	\country{Australia}
}
\author{Dali Kaafar}
\email{dali.kaafar@mq.edu.au}
\orcid{0000-0003-2714-0276}
\affiliation{%
	\institution{Macquarie University}
	\city{Sydney}
	\country{Australia}
}

\renewcommand{\shortauthors}{Carpentier et al.}

\begin{abstract}
Interactions with online Large Language Models raise privacy issues where providers can gather sensitive information about users and their companies from the prompts.
While textual prompts can be sanitized using Differential Privacy, we show that it is difficult to anticipate the performance of an LLM on such sanitized prompt. Poor performance has clear monetary consequences for LLM services charging on a pay-per-use model as well as great amount of computing resources wasted.
To this end, we propose a middleware architecture leveraging a Small Language Model to predict the utility of a given sanitized prompt before it is sent to the LLM. We experimented on a summarization task and a translation task to show that our architecture helps prevent such resource waste for up to 20\% of the prompts.
During our study, we also reproduced experiments from one of the most cited paper on text sanitization using DP and show that a potential performance-driven implementation choice dramatically changes the output while not being explicitly acknowledged in the paper.
\end{abstract}

\keywords{Privacy-Utility Trade-off, Privacy-Aware Resource Efficiency, Resource Saving, Small Language Model, dx-privacy}


\maketitle

\section{Introduction}
Large Language Models (LLMs) are now being used for a variety of tasks such as document retrieval, code explanation, document summarization, and image generation. 
The largest models, with respect to the number of parameters, currently yield the strongest performance~\cite{wei2022Emergent}. These large models require significant computing resources for training, which only a few entities worldwide possess the capability to provide. Even after the training phase, querying the largest models remains resource-intensive and beyond the capabilities of an average personal computer. LLMs are then typically hosted by these same entities and made available online as a service (either free or paid) that users query (or prompt) and receive a response\footnote{e.g., \url{https://chatgpt.com/} or \url{https://mistral.ai/}}. During this process, any sensitive information contained within the prompt is received by the service provider.

Sharing information through LLM prompts raises privacy concerns in both professional and personal settings. For instance, trade secrets\footnote{\href{https://www.techradar.com/news/samsung-workers-leaked-company-secrets-by-using-chatgpt}{techradar.com/news/samsung-workers-leaked-company-secrets-by-using-chatgpt}} or client data \cite{cyberhaven2023ChatGPT} can be exposed.
In addition to data being revealed to the provider itself, prompts may also be used to improve LLMs, which in turn are known to memorize their training data and may disclose them when subsequently queried by other users~\cite{carlini2019secret}. 

While textual prompts can be manually redacted and sanitized by human experts, such process is time consuming and often not practical in several applications \cite{gordon2013mra}.
Automatic sanitization is generally carried out either using Differential Privacy (DP) applied on text \cite{feyisetan2020Privacy, yue2021Differential, asghar2024dxprivacy, qu2021Natural}, Generalization \cite{hassan2023Utilitypreserving, olstadGenerationSelectionReplacement2023} or rewriting assisted by language models \cite{weggenmann2022DPVAE, igamberdiev2023DPBART}. It can be applied on the entire text or on specific, privacy-exposing elements detected by Named Entity Recognition~\cite{chen2023Hide, chong2024Casper, papadopoulou2023Neural}.

All these sanitization techniques involve altering the text to various degrees which can reduce its utility~\cite{zhang2024No}. If said text is a prompt intended to be sent to an LLM for a task (e.g., summarization), the result might suffer in correctness, accuracy and/or coherence compared to what the LLM would have produced with the original prompt. Excessive alteration can even lead to totally useless results which are wasted resources for the provider and unnecessary expenses for the user when the LLM usage incurs a fee. Moreover, we show that it is difficult for a user to anticipate the performance of an LLM on a sanitized prompt, making it challenging to avoid wasting resources. 
Throughout this work, \emph{utility} denotes the degree to which an input (e.g., a prompt) enables high-quality results in a particular computation (e.g., a language model task), as measured by an appropriate evaluation metric (e.g., accuracy or semantic similarity). This aligns with standard usage in the privacy literature, as seen in~\cite{zhang2024No,feyisetan2020Privacy,qu2021Natural,yue2021Differential,wood2018Differential}.

In this paper, we address the following problem of practical concern. To prevent employees from inadvertently submitting sensitive or Personally Identifiable Information (PII) into prompts, a company aims to sanitize the prompts before they are sent to the LLM provider. 
However, an excessive loss in utility caused by the sanitization is undesirable from the company's perspective as it incurs costs for every prompt sent to the LLM provider.
In this context, we investigated the possibility of locally predicting the LLM's performance on a given sanitized prompt. This way, we can avoid sending excessively sanitized prompts which will produce meaningless results and waste resource. Our prediction is carried out by a trusted middleware positioned between the employee and the online LLM, for example on the employee’s device or within the company's infrastructure.

The middleware leverages a Small Language Model (SLM) of manageable size for the company ($\leq$1B parameters in our tests). While less powerful than LLMs, the SLM improves prediction accuracy and aids in deciding whether to use the online LLM with privacy protection or rely on the trusted SLM.
In fact, if the sanitized prompt is predicted to lead to poor performance, it is not sent to the LLM, avoiding resource waste, and the SLM handles the task itself using the original prompt. This approach aims to offer the company and its employees the best value for their money by balancing privacy, utility, and financial resources.

We experimented on text sanitization using DP \cite{feyisetan2020Privacy, qu2021Natural, asghar2024dxprivacy, yue2021Differential} as 
DP is currently the prevailing framework for privacy-respecting computations. We focus on two popular tasks: a summarization task and a machine translation task, testing five SLMs and two LLMs. We show that leveraging a local SLM enables to save the computing resources of up to 20\% of the prompts.

Overall, in this paper we make the following contributions:
\begin{enumerate}
	\item We propose a middleware architecture which locally evaluates the quality of a prompt sanitized using a privacy-preserving text sanitization method. This evaluation occurs before sending the prompt to an LLM, aiming to prevent resource waste if the sanitization has degraded the prompt's utility to the point where satisfactory results are unlikely. We refer to this architecture as the \emph{utility assessor}.
	\item We empirically evaluate the utility of prompts sanitized by $\dx$-privacy, a word-level Differential Privacy method leveraging the Multidimensional Laplace Mechanism.
	Our evaluation targets a summarization task and a translation task performed on a publicly available dataset, leveraging five SLMs and two LLMs. We show the colossal impact of sanitization where none of the models are able to produce results of any utility for amounts of noise usually considered non-private in ordinary DP (e.g., $\varepsilon=35)$. 
	With lower amounts of noise, the same $\varepsilon$ value may provide no utility when applied on one prompt while offering very high utility when applied on another as word-level DP can alter semantics to varying degrees. Hence, predicting the utility of sanitized prompts requires an advanced technique such as our utility assessor.
	
	\item In order to better predict the utility of a prompt---whether it would waste resources or not---we rely on several numerical features beyond the $\varepsilon$ value, available through our utility assessor. This includes semantic similarities involving the original prompt and the SLM's results on the original and sanitized prompts.
	Using these features, we trained a regressor to predict the performance of an LLM on a given sanitized prompt. Compared to the use of $\varepsilon$ alone, we show that using the utility assessor's features enables saving the computing and monetary resources of up to 20\% of the prompts that were previously wasted.
	
	\item We identify an issue in Feyisetan et al.~\cite{feyisetan2020Privacy} which, to the best of our knowledge, is one of the most cited paper on Differential Privacy applied on text through the Multidimensional Laplace Mechanism. By replicating some of their experiments, we show that the potential reliance on a approximate nearest neighbor search in multidimensional vector spaces ---while understandable for performance--- completely changes the results of the experiments while not being explicitly mentioned in their paper. Instead, our implementation rigorously following their definition (i.e., an \emph{exact} nearest neighbor search) outputs the \emph{original input} 99\% of the time for most of our replicated experiments, not providing any privacy protection.
\end{enumerate}

The code for all of our experiments is available online\footnote{\url{https://github.com/r-carpentier/Preempting-Text-Sanitization-Utility}\label{footnoteCodeRepo}}. 
The paper is organized as follows: \cref{sec:backDX} provides background knowledge about DP-based text sanitization. \cref{sec:archi} motivates the problem and describes our middleware architecture (first contribution). In \cref{sec:settings} we describe our experiment settings and detail an issue we faced during our initial implementation of the sanitization mechanism (fourth contribution). \cref{sec:expe} we experimentally validate the benefits of our middleware architecture (second and third contribution). \cref{sec:sota} presents the related works on text sanitization and the privacy of language models' prompts. Finally, we discuss the limits of our architecture in \cref{sec:limitations}.

\section{Background on Differential Privacy for Text}\label{sec:backDX}
In this section we provide necessary background information about differential privacy, how concepts of differential privacy can be applied in the text domain, and the specific technique for text sanitization used in our work.

\subsection[Differential Privacy and dx-Privacy]{Differential Privacy and $\dx$-privacy}
Differential privacy (DP) is a formal framework providing privacy in data analysis. Informally, an algorithm handling a set of data is said to be differentially private if its output is not significantly affected by the presence or absence of one particular input~\cite{dwork2014algorithmic}.
DP is achieved by adding properly calibrated noise to the result of the computation itself or to the input data in the case of Local DP \cite{kasiviswanathanWhatCanWe2011}.
While DP is usually applied on structured data (numbers, tabular etc.), applications on unstructured data have also been proposed \cite{zhao2022survey} in particular on text \cite{feyisetan2020Privacy,yue2021Differential,qu2021Natural, andres2013Geoindistinguishability, asghar2024dxprivacy}.  
In this paper, we focus on DP as our text sanitization method because differential privacy is currently regarded as the gold standard for privacy-preserving data publication \cite{balle2020hypothesis}. Also, the privacy guarantees of these methods hold for the entire text as they do not rely on the performance of a PII detector selecting entities to sanitize. See \cref{sec:sota} for an overview of the state of the art in general text sanitization techniques. 

For this work we leverage $\dx$-privacy, a relaxation of Local DP where the domain of input data is evaluated by an arbitrary distance metric $d$ \cite{chatzikokolakis2013Broadening}. In Local DP, the indistinguishability property applies equally to all elements of the input domain. This means that given a certain output, the mechanism must make it equally plausible that any input could have produced said output. Conversely, $\dx$-privacy supports indistinguishability relative to $d$, where similar inputs will produce more similar outputs, making said inputs distinguishable from other dissimilar inputs. 
Overall, $\dx$-privacy is designed to improve the utility over ordinary DP by acknowledging some information about the input in the output.
An example of $\dx$-privacy is given in \cite{andres2013Geoindistinguishability}, called Geo-indistinguishability, which avoids disclosure of a user's exact geographic location by releasing a differentially private substitute. In this example, the metric used is the Euclidean Distance with closer locations favored over more distant ones from the original. The radius of what is considered close depends on the privacy parameter $\varepsilon$.

\subsection{Text Sanitization with DP - Theory}

Applying $\dx$-privacy on textual data is made possible by the emergence of vector representations of text including word2vec~\cite{word2vec}, GloVe~\cite{pennington2014GloVe} and the inner mechanism of language models \cite{devlin2019BERT}. Vast quantities of text are used to progressively learn the representation of tokens (i.e., a word or a word segment) as $n$-dimensional vectors called embeddings where the distances in the embedding space express the semantic difference between tokens.

Sanitization of a text is then performed by adding differentially private noise to the embeddings of its tokens, for example by sampling noise via the Multidimensional Laplace Mechanism~\cite{feyisetan2020Privacy}. Formally, assume a set of tokens $\mathcal{X}$ and a token embedding model $\embmod:\mathcal{X}\rightarrow\mathbb{R}^n$ associating each token with an $n$-dimensional embedding vector. $\mathcal{X}$ is equipped with a distance function $d:\mathcal{X}\times \mathcal{X}\rightarrow \mathbb{R}^{+}$ which is a metric with $d(x_1, x_2)=\lVert \embmod(x_1)-\embmod(x_2)\rVert$. Assume a privacy parameter $\varepsilon\in \mathbb{R}^*_+$. A randomized mechanism $M: \mathcal{X}\rightarrow\mathcal{Y}$ is said to be $\varepsilon \dx$-private~\cite{feyisetan2020Privacy} if for any input $x_1, x_2 \in \mathcal{X}$ and any output $y\in \mathcal{Y}$:
\begin{equation*}
	\frac{Pr[M(x_1)=y]}{Pr[M(x_2)=y]}\leq e^{\varepsilon d(x_1, x_2)}
\end{equation*}
Intuitively, consider the case of text-to-text sanitization where the output domain is also tokens and thus $\mathcal{Y}=\mathcal{X}$. The closer two tokens $x_1$ and $x_2$ are in the embedding space according to the distance metric, the more similar the probabilities are that they will be mapped to the same output token $y$ by the mechanism. This means that tokens with similar meanings (i.e., close embeddings) are nearly indistinguishable from the perspective of the output distribution of the mechanism as they are almost equally likely to produce any given output token $y$. Note that Local DP is a special case of $\dx$-privacy where all distances are equal, i.e. $\forall x_1, x_2\in \mathcal{X}: x_1\neq x_2 \Rightarrow d(x_1, x_2) = 1$.

\subsection{Text Sanitization with DP - Application}
\label{subsec:santize}
\begin{algorithm}[b]
	\caption{$\dx$-privacy}\label{alg:dx}
	\begin{flushleft}
		\textbf{Input:} Token $x\in \mathcal{X}$, $\varepsilon\in \mathbb{R}^*_+$\\
		\textbf{Output:} Sanitized Token $x'\in \mathcal{X}$
	\end{flushleft}
	\begin{algorithmic}[1]
		\STATE $\noise = mv\_normal(\mu=\mathbf{0}_n, \Sigma=\mathbf{1}_n)$      \RIGHTCOMMENT{Sample multiv. normal dist.}
		\STATE $\noise = \noise/\norm{\noise}$ \RIGHTCOMMENT{Normalize}
		\STATE $\noise = gamma(k=n, \theta=1/\varepsilon) * \noise$\RIGHTCOMMENT{Sample gamma dist.}
		\STATE $e' = \embmod(x) + \noise$      \RIGHTCOMMENT{Add noise to the embedding of $x$}
		\STATE $e = argmin_{z\in \mathcal{X}}\norm{\embmod(z)-e'}$ \RIGHTCOMMENT{Find nearest neighbor in $\mathcal{X}$}
		\STATE $d_{NN}(e, z)=i$ if $z\in \mathcal{X}$ is the $i$th nearest neighbor of $e$
		\STATE Sample $x'\in \mathcal{X}$ with $Pr(x') \propto \exp(-\varepsilon\cdot d_{NN}(e, x'))$
		\STATE \textbf{return} $x'$
	\end{algorithmic}
\end{algorithm}
Text-to-text sanitization satisfying $\dx$-privacy takes many forms \cite{feyisetan2020Privacy, qu2021Natural, yue2021Differential}, the specific technique of our work follows that of Feyisetan et al.~\cite{feyisetan2020Privacy} with additional steps proposed by Asghar et al.~\cite{asghar2024dxprivacy}. 
\cref{alg:dx} outlines the sanitization process of a given token $x\in \mathcal{X}$ considering a privacy parameter $\varepsilon$. Lines \textit{1} to \textit{3} detail how to sample the noise vector $\eta$: First, the direction of the noise is chosen via a sample from the multivariate normal distribution of dimension $n$, with a zero mean and the identity matrix as the covariance (Line \textit{1}). It is then normalized (Line \textit{2}). On line \textit{3}, the magnitude of the noise is sampled with a gamma distribution of shape $n$ and scale $1/\varepsilon$. By multiplying it by the result of line \textit{2}, we now have a noise vector $\noise$ with a random direction and a magnitude scaled by $\varepsilon$.
By adding the noise to the embedding of $x$ on line \textit{4}, we obtain $e'$ which is a point of $\mathbb{R}^n$ that with high probability does not correspond to any word in $\mathcal{X}$. On line \textit{5} we then perform a nearest neighbor search to find the word $z\in \mathcal{X}$ where $\norm{\embmod(z)-e'}$ is minimized.

While the mechanism by Feyisetan et al.~\cite{feyisetan2020Privacy} considers $e$ (line \textit{5}) as the replacement token, the work by Asghar et al.~\cite{asghar2024dxprivacy} introduces additional steps to ensure enough close neighbors of the input token $x$ can be produced as output (depending on $\varepsilon$). This is a desirable property as close neighbors of a token in the embedding space are semantically-related, and a property that Asghar et al.~\cite{asghar2024dxprivacy} argues the original mechanism does not support due to the behavior of the nearest neighbor search of the noisy embedding $e'$ (line \textit{5}) in highly-dimensional embedding spaces.
The additional steps are outlined in lines \textit{6} and \textit{7}: we first rank all the elements in $\mathcal{X}$ based on their distance to $e$ and then sample a replacement token from $\mathcal{X}$ where the probability of selecting each element is proportional to an exponential of its rank in the neighbor list of $e$ multiplied by $-\varepsilon$.

For a text containing several tokens, the mechanism is applied independently to each token. 
It is important to note that \cref{alg:dx} might output $x' = x$. Indeed, $x'$ is necessarily in the neighbor list of $e$ and gets assigned a non-zero probability of being sampled on line \textit{7}. In this case the token $x$ is not modified by the mechanism.
Under DP, the $\varepsilon$ parameter governs the privacy-utility trade-off: higher $\varepsilon$ values introduce less noise, reducing privacy protection. In $\dx$-privacy however, $\varepsilon$ also depends on the sensitivity of the distance metric $d$, meaning its interpretation varies across different metrics. As a result, $\varepsilon$ values in $\dx$-privacy are not directly comparable to those in ordinary DP, and their absolute values in this paper do not hold the same significance as the ones found in papers using DP on numerical data.
Furthermore, the meaning of $\varepsilon$ also varies with the structure of the chosen embedding model $\embmod$ \cite{xu2021DensityAware}. Indeed, depending on the distance between elements in the embedding space, the same noise magnitude (e.g., $\varepsilon=30$ in line 3 of \cref{alg:dx}) can be relatively large (i.e. reaching distant neighbors) in one embedding model while being small in another.

Thus, choosing the value for $\varepsilon$ is not straightforward for end users \cite{nanayakkara2023What} as we will illustrate in the next section. 
As such, in this paper we focus on $\dx$-privacy as our main sanitization method and propose an architecture to assess degrees of $\dx$-privacy's impact on the utility of prompts in an attempt for end-users to better navigate the trade-off between privacy and utility and avoid resource waste.

Additionally, in \cref{sec:expe2} we reproduce some experiments from Feyisetan et al.~\cite{feyisetan2020Privacy} on which \cref{alg:dx} is based. We show that their implementation potentially relies on an approximate nearest neighbor search for line \textit{5} of \cref{alg:dx} while not explicitly acknowledging it in the paper. While understandable for the sake of performance, we demonstrate this choice to have an enormous impact on the output of the mechanism.
Instead, an implementation rigorously following their definition (i.e., an exact nearest neighbor search) will output the input token 99\% of the time for most of our replicated experiments, not providing any privacy protection. Asghar et al.~\cite{asghar2024dxprivacy} later formally studied the behavior of the mechanism and proposed a fix that we used in line \textit{6} and \textit{7} of \cref{alg:dx}.

\section{Middleware Proposal}\label{sec:archi}

In this section we motivate the need for more sophisticated means of assessing the privacy-utility tradeoff in prompt sanitization, followed by an overview of our proposed middleware architecture. 

\subsection{Utility and Privacy Trade-off for Prompts}\label{subsec:archi_context}
When querying online LLMs, a handful of tasks require the submission of an instruction accompanied by a text on which the task is performed. For example, a summarization task may consist in an instruction 
"Summarize the following text" and a text. Other tasks follow the same model, such as Question Answering where the LLM will answer a question using information contained within a textual document, or Data Extraction where specific information are extracted from a document. In these cases, the text to be processed might contain sensitive elements such as PII or trade secrets that the user or its employer does not want to be revealed to the LLM's provider.
For this reason, the user (or its employer on the user's behalf) will resort to sanitizing the associated text. For clarity, in the rest of this paper we refer to this associated text as the \emph{prompt} as we consider that the associated instruction does not require privacy protection.  

Sanitizing the prompt can be achieved with manual sanitization at a heavy burden and cost to users~\cite{gordon2013mra}, or one can leverage automatic methods. In particular, we consider the differentially-private text sanitization method described in \cref{subsec:santize} (See \cref{sec:sota} for an overview of other sanitization approaches). Because sanitization modifies the content of the prompt, it affects the performance of the LLM task performed on said prompt. This is referred to as the \emph{utility} of the prompt and is typically measured by an appropriate evaluation metric (e.g., accuracy or semantic similarity) following standard usage in the privacy literature, e.g.~\cite{zhang2024No,feyisetan2020Privacy,qu2021Natural,yue2021Differential,wood2018Differential}.

\subsubsection{Example of Sanitized Text} To illustrate the extent to which a text is altered by differentially-private text sanitization, consider the following text sample:
\begin{textsample}
	Emily Carter, born on April 12, 1990, resides at 482 Maple Street, Springfield, IL, and her Social Security Number is 123-45-6789. Her credit card number, 4111-1111-1111-1111, expires in 06/27, and her personal email is emily.carter90@email.com.
\end{textsample}
Sanitizing this text using \cref{alg:dx} with $\varepsilon=100$ results in the following text\footnote{We use the token embedding model $\embmod$ of the bart-large-cnn language model.}:

\begin{textsample}
	Rachel guiName,Born inApril 13, 1990. resideAt 482 Maple Street in Springfield,, andHer social Securitynumber has 127.456789, Her credit card Number. 6109 and-reportprint,1111, expired in 06/23. and her personal email was em.carter60@emailcom
\end{textsample}
\noindent Note that this is the exact output: we did not omit spaces or add characters during the writing of this paper. We notice that the result lacks proper grammar and punctuation. Also, while the text conserves its meaning to some extent (i.e., presenting a person), some information is partially altered such as the social security number, while others like the name are entirely modified. 

Decreasing $\varepsilon$ achieves better privacy but also lowers utility. Below is a sanitized version\footnote{Included as an image to avoid formatting issues} of the text using $\varepsilon=35$:
\begin{textsample}
	\noindent\includegraphics[width=\columnwidth]{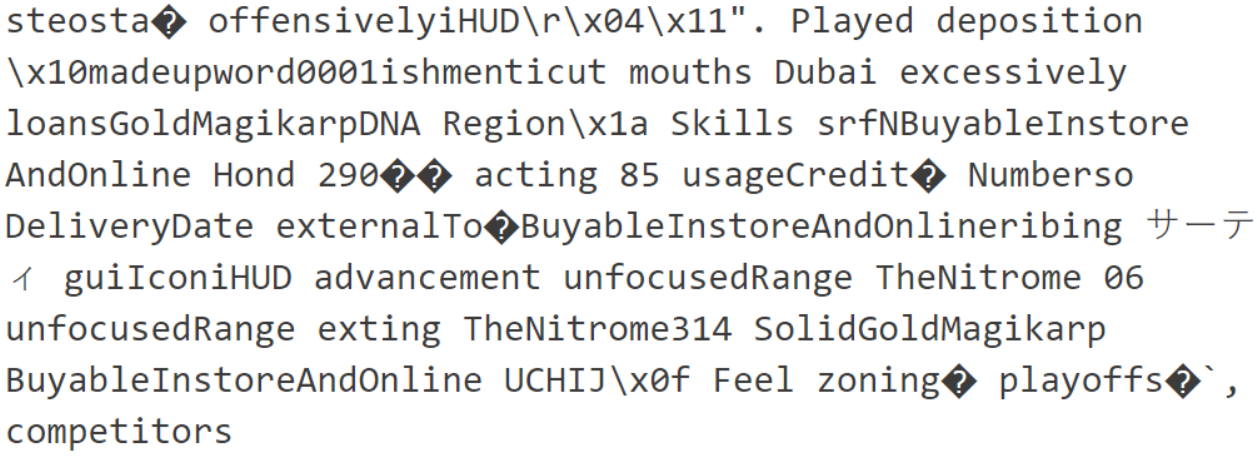}
\end{textsample}
\noindent The text has lost all of its meaning, grammar and punctuation. Some characters are not even English and others cannot be printed.

Such text deterioration is expected given \cref{alg:dx}. Indeed, considering a token as a point of an $n$-dimensional vector space, the noise added by the mechanism can point towards any direction, with its magnitude scaled by $\varepsilon$. Consequently, a sufficiently low value of $\varepsilon$ potentially leads to a perturbed point relatively far from the original point.
Because embedding models are trained on vast amounts of textual data, they contain a highly diverse set of tokens that can be chosen as output by the mechanism, hence the non-printable and non-English characters.

\subsubsection{Performance Consideration}\label{sec:sanit_perf_consideration} Any language model task computed on such sanitized prompt will suffer in performance. A user looking for satisfactory performance of an LLM on such task might consider a simple yet insecure solution: sending the same prompt multiple times, each time decreasing the sanitization degree. This can be repeated until the desired result quality is achieved or the privacy level becomes unacceptably low. First, this approach is inefficient in terms of computing and financial resources as only the last query result is useful to the user. Second, it poses privacy risks: As \cite{dwork2008differential} puts it simply, repeated sanitized queries of the same prompt can be averaged, effectively canceling the noise out and revealing the original input.

Thus, the user is in need of a solution to anticipate the performance of the LLM before actually sending the prompt. Without sanitization, such performance is usually influenced by the exact task and the LLM used for the task. The sanitization introduces another factor, mainly driven by $\varepsilon$. However, as mentioned in \cref{sec:backDX}, $\varepsilon$ values are not similar to ordinary DP making it more challenging to select an appropriate value. For example, the last text sample above is sanitized using $\varepsilon=35$, a value that would be considered non-private in ordinary DP, while it is still clearly too private to get any utility. $\varepsilon$ also depends on the specific embedding model used (See \cref{sec:backDX}). Moreover, the same $\varepsilon$ value can affect different texts to various degrees as we will see in \cref{sec:expe}. 
Overall, predicting the performance of an LLM on a given sanitized prompt before it is sent is challenging.

\subsection{Utility Assessor Architecture}
We propose a solution architecture where a user benefits from a trusted middleware including an SLM to assess the utility of a given sanitization of a prompt. As the prompt is going to be sent to an LLM for inference, this assessment enables to anticipate the LLM's performance and avoid wasting its resources if the expected result will not meet a desired quality. 
Intuitively, by analyzing the relationship between the original and sanitized prompts, as well as the SLM's performance on both, we believe the middleware is able to more accurately predict how the LLM will perform on the sanitized prompt.

\begin{figure}
	\centering
	\includegraphics[clip, trim=0 10.5cm 19.1cm 0, width=\linewidth]{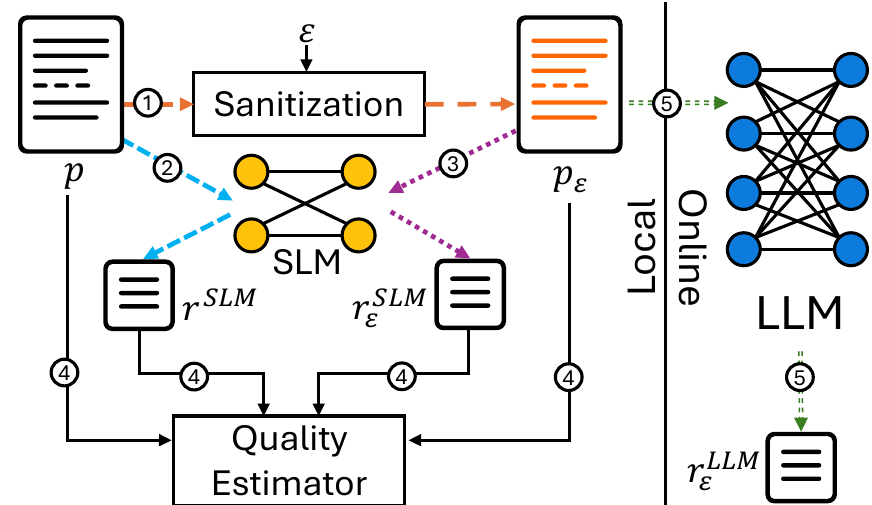}
	\caption{Utility Assessor Architecture. $p$ represents the original prompt. $p_\varepsilon$ denotes the sanitized version of the prompt. $r^{SLM}$ and $r^{SLM}_\varepsilon$ designate the SLM's result on the original and sanitized prompts, respectively. $r^{LLM}_\varepsilon$ identifies the LLM's result on the sanitized prompt.}
	\label{fig:architecture}
\end{figure}

\subsubsection{Components} The architecture we consider is illustrated in \cref{fig:architecture}. It deals with language model tasks where the prompt $p$ might contain sensitive elements. Locally (left-hand side of the figure), the user can apply a \emph{Sanitization} procedure to $p$, parametrized by $\varepsilon$, to obtain the sanitized prompt $p_\varepsilon$ (Step 1). They have access to an SLM to run the language model task on $p$ and on $p_\varepsilon$ to obtain the associated results $r^{SLM}$ and $r^{SLM}_\varepsilon$ (Step 2 and 3). They also have a \emph{Quality Estimator} procedure which evaluates the quality of the SLM's results (Step 4).
Finally, when a given $p_\varepsilon$ is assessed to satisfy the quality criterion, it is sent to the LLM hosted online (right-hand side of the figure) to perform the language model task (Step 5). 
Otherwise if the desired quality is not met, the LLM is not leveraged to preserve resources, but the user can nonetheless benefit from $r^{SLM}$ as the result to their prompt.

We define an SLM as a model capable of running on the edge, either on the user's own machine or on a machine hosted by their company. The large size of online LLMs, such as ChatGPT-4 with an estimated 1.8 trillion parameters\footnote{\url{https://the-decoder.com/gpt-4-architecture-datasets-costs-and-more-leaked/}}, makes them impractical for local hosting.
In contrast, we consider SLMs with at most one billion parameters (such model with 32-bit precision necessitates approximately 4GB of VRAM). Thus, their execution is considered cheap in terms of computing resources. They can be either general-purpose or fine-tuned for a specific task. Fine-tuned SLMs might have better performances on a task, and thus might improve the quality of the prediction of the LLM's performance on the same task. However, they are not task-agnostic and several SLMs of this type might be needed to cover a full range of language model tasks. We tested both types in our experiments presented in \cref{sec:expe}.

The Quality Estimator is task-dependent and judges the quality of $p_\varepsilon$, $r^{SLM}$ and $r^{SLM}_\varepsilon$. For example, semantic similarity between $p$ and $r^{SLM}_\varepsilon$ can be used to assess the proximity of a text and its summary in a summarization task. In a text continuation task, we can evaluate $r^{SLM}$ and $r^{SLM}_\varepsilon$ with coherence or diversity. Similarities, coherence and any other quality metric are considered as features and are subsequently used in a regression task to predict the utility of $r^{LLM}_\varepsilon$. Such regressor can be trained on the flow: while sanitized prompts are being sent to the LLM and answers are received, input features can be extracted from $p_\varepsilon$, $r^{SLM}$ and $r^{SLM}_\varepsilon$, while the target feature can be derived from $r^{LLM}_\varepsilon$.
In particular, we experimented with a regression-based prediction as a simple and inexpensive process. Conversely, fine-tuning a language model for the same purpose requires more computing resources and knowledge (e.g. about hyperparameters tuning). 
Limitations of the Quality Estimator, especially task support, are discussed in \cref{sec:limitations}.

\subsubsection{Threat Model and Problem Statement} The provider of the online LLM is honest but curious. They deliver the service but collects data from received prompts. The user's local environment as well as the communication between the user and the online LLM are considered trusted.
Considering the utility assessor architecture, the computing model and the threat model defined above, can we predict the performance of an LLM?
Put simply, can a trusted, smaller language model help users balance privacy and utility when prompting a large and untrusted language model?

\section{Experimental Setup}\label{sec:settings}
In this section, we outline the settings of our experiments including the dataset and language models used. We also expose an issue we faced during our initial tests with the sanitization mechanism. Results are presented in \cref{sec:expe}.

\subsection{Tasks, Dataset and Models}

\textbf{Tasks.} In our experiments, we target a summarization task and a machine translation task. 
In the summarization task, an accurate summary of an English text has to be produced by the language model in a limited number of tokens. For machine translation, the model has to provide a proper translation of an English text into French. We chose these particular tasks as they are popular document-driven language model task and are notably relevant in a work setting: summarizing long e-mails and reports or translating internal documents in a multinational company are desirable features for employees. In these cases, the company itself would want to impose privacy protection as these documents may contain confidential information or PIIs. Translation to French was specifically chosen because one of the authors can understand the language.

\noindent\textbf{Dataset.}
We use the Multi News dataset \cite{fabbri2019Multinewsa} which contains news articles from multiple sources.
We concatenated its train, validation and test subsets and cleaned it to the best of our ability by removing unrelated headers, faulty entries, and other abnormalities (See our code release\hyperref[footnoteCodeRepo]{\textsuperscript{\ref{footnoteCodeRepo}}} for full details of data preparation).
As some of our SLMs only support inputs of at most 1024 tokens, for a fair comparison we only select texts of this length, with the final evaluation dataset containing 11,121 documents. For all experiment we randomly sample 1,500 elements from this dataset.

\noindent\textbf{SLMs.} We use three different SLMs to summarize text: T5 Small fine-tuned for summarization~\cite{falconsai2023t5} (60M parameters), bart-large-cnn~\cite{lewis-etal-2020-bart} (400M parameters) and Llama3.2-1B-Instruct~\cite{grattafiori2024llama3herdmodels} (1B parameters). 
Likewise, for translation to French we use three models: T5 Small~\cite{raffel2020t5} (60M parameters), Opus-MT~\cite{tiedemann-thottingal-2020-opus,tiedemann-2020-tatoeba} (230M parameters) and the Llama model mentionned above. These models were chosen for their popularity on the huggingface platform\footnote{They have respectively 1M, 100M, 15M, 143M and 200K all-time downloads.}
and because they exhibit diverse sizes while still fitting our SLM definition (i.e. $\leq$ 1B parameters).
Llama3.2-1B is a general-purpose model while the others are mostly limited to a single task. 

\noindent\textbf{LLMs.} For the target LLM to be queried we tested two models: Llama-3-8B-Instruct~\cite{grattafiori2024llama3herdmodels} and Gemini-2.0-flash~\cite{geminiteam2025geminifamilyhighlycapable}. We choose to host the first model ourselves to reduce experiment costs and enhance reproducibility, as online service offerings evolve over time. While Llama-3 is smaller than trillion-parameter models available online, the size gap between Llama-3 and our SLMs remains significant: T5, Opus-MT, BART, and Llama3.2-1B are respectively 132x, 34x, 20x, and 7x smaller in terms of parameter count.
We also include Gemini to involve a recent online large language model, though the exact number of parameters for Gemini-2.0-Flash is unknown.

\noindent\textbf{Task Configuration.} For summarization, we requested models to limit their output to 142 tokens as it is the default value for the bart-large-cnn model and we found it to be a suitable choice for all models. For translation, outputs are set to be at most 30\% longer than the input. General-purpose models were driven to summarize or translate using specific prompts detailed in the code release\hyperref[footnoteCodeRepo]{\textsuperscript{\ref{footnoteCodeRepo}}}. 

\noindent\textbf{Quality Estimator.} We assess summary quality as the similarity between a prompt and its summary and measured by the all-mpnet-base-v2 model~\cite{reimers-2019-sentence-bert}. Likewise, translation quality is measured as the similarity between a prompt and its translation using the paraphrase-multilingual-mpnet-base-v2~\cite{reimers-2020-multilingual-sentence-bert}. These models generate one embedding vector for an entire text, and the similarity between two texts is measured as the cosine similarity between their embeddings. Similarities range from -1 to 1: -1 indicates opposing texts, 0 suggests no correlation (orthogonal texts), and values close to 1 signify high similarity. 
While we have also experimented with other text similarity models, namely gte-large-en-v1.5~\cite{alibaba2024gteLarge,zhang2024mgtegeneralizedlongcontexttext} and modernbert-embed-base~\cite{nomicai2024moernbertembed, nussbaum2024nomic}, the result were not satisfactory as they designate overly sanitized texts (i.e., $\varepsilon=1$) as similar to genuine English text. We have also experimented with a translation quality estimator model named wmt22-cometkiwi-da~\cite{unbabel2023wmt22, rei-etal-2022-cometkiwi} which does not rely on embedding similarity, but we found that the model does not perform well on texts longer than a few sentences.

\subsection{Issues with the Sanitization Mechanism}\label{sec:expe2}
We sanitize the texts from the Multi News dataset using the mechanism based on the work of Feyisetan et al.~\cite{feyisetan2020Privacy} and detailed in \cref{alg:dx}. Our initial implementation of the mechanism did not produce the same results reported in Feyisetan et al.~\cite{feyisetan2020Privacy}. In this section we detail the issue we faced and the specific configuration we used in the rest of the paper.

To expose the issue, we replicate two experiments presented in Feyisetan et al~\cite{feyisetan2020Privacy} as faithfully as possible. Note that while the original paper does not contain a reference to a codebase, a repository\footnote{\url{https://github.com/awslabs/sagemaker-privacy-for-nlp}} by the same institution claims to use the same algorithm and cites the paper itself. We do not assume this code to be the exact same used in Feyisetan et al.~\cite{feyisetan2020Privacy}, but we examined the repository to decrease the likelihood of an implementation error on our side for the $\dx$-privacy mechanism. A notable difference we observe in this codebase is the employment of a nearest neighbor search (line 5 of \cref{alg:dx}) using an Approximate Nearest Neighbor (ANN) library~\cite{spotifyAnnoy}. This is in contrast with the algorithm presented in the original paper~\cite[Algorithm 1]{feyisetan2020Privacy} which suggests an Exact Nearest Neighbor (ENN). We tested both variations in our experiment replication detailed below and included in our code release\hyperref[footnoteCodeRepo]{\textsuperscript{\ref{footnoteCodeRepo}}}. For precision, in this section the mechanism ends at line \textit{5} of \cref{alg:dx} and considers $e$ as the replacement token.

\subsubsection{Example 1: Individual Words}
\begin{table*}
	\centering
	\begin{tabular}{|c|c|c||c|c||c|c|}
		\hline
		\multirow{2}{*}{$\varepsilon$} & \multicolumn{2}{c||}{`encryption'} & \multicolumn{2}{c||}{`hockey'} & \multicolumn{2}{c|}{`spacecraft'} \\ \cline{2-7} 
		& \textbf{ANN} & \textbf{ENN}      & \textbf{ANN} & \textbf{ENN}  & \textbf{ANN} & \textbf{ENN}  \\ \hline
		19  & (315) encryption  & (995) encryption & (593) hockey    & (996) hockey    & (385) spacecraft & (985) spacecraft\\ 
		& (39) encrypted    & (1) slosberg     & (35) nhl        & (1) cellspacing & (54) spaceship   & (2) nasa        \\
		& (26) cryptographic& (1) snubbing     & (13) soccer     & (1) canada      & (25) orbit       & (2) orbit        \\
		\hline
		
		25  & (550) encryption  & (1000) encryption & (797) hockey  & (999) hockey & (590) spacecraft  & (999) spacecraft   \\ 
		& (63) encrypted    &                   & (33) nhl      & (1) played    & (71) spaceship   & (1) spaceship  \\
		& (26) cryptographic&                   & (9) soccer    &               & (19) satellites  &  \\
		\hline   
		
		35  & (822) encryption  & (1000) encryption & (943) hockey & (1000) hockey & (788) spacecraft  & (1000) spacecraft \\ 
		& (65) encrypted    &                   & (11) nhl     &               & (79) spaceship  &  \\
		& (13) cryptography&                    & (8)  soccer  &               & (17) orbiter  &  \\
		\hline  
		
		43  & (910) encryption  & (1000) encryption & (989) hockey   & (1000) hockey   & (876) spacecraft  & (1000) spacecraft  \\ 
		& (36) encrypted    &                   & (5) nhl        &                  & (56) spaceship &   \\
		& (17) cryptographic&                   & (1) coyotes   &                  & (8) orbiter  &  \\
		\hline   
	\end{tabular}
	
	\caption{Frequency of the most common output words generated by 1000 sanitizations of \emph{encryption}, \emph{hockey} and \emph{spacecraft} from the GloVe embedding model using Approximate Nearest Neighbor (ANN) and Exact Nearest Neighbor (ENN) methods.} 
	\label{tab:dxexpe1}
\end{table*}
In one of the experiments, the authors use $\dx$-privacy to sanitize individual words from the word embedding model GloVe~\cite{pennington2014GloVe}\footnote{While several versions of GloVe have been released, Feyisetan et al.~\cite{feyisetan2020Privacy} mention that it contains 300d and 400,000 words which corresponds to glove.6B.300d.pkl.}. A word is sanitized 1,000 times with different values of $\varepsilon$ and examples of output are provided for the words \emph{encryption}, \emph{hockey} and \emph{spacecraft}.
We reproduce this experiment and present our findings in \cref{tab:dxexpe1} which contains the top three output words and their frequencies when sanitizing with different values of $\varepsilon$
\footnote{
	\cite[Table 1]{feyisetan2020Privacy} does not display $\varepsilon$ values but average $N_w$ values, where $N_w$ is the empirical frequency of the input word being returned by the mechanism.
	We infer the corresponding $\varepsilon$ values used by the authors based on \cite[Figure 3(b)]{feyisetan2020Privacy} which plots average $N_w$ values depending on $\varepsilon$. 
}
for the two nearest neighbor methods. 
First, we observe that with ANN we obtain results similar to \cite[Table 1]{feyisetan2020Privacy} where words such as \emph{encrypted} and \emph{cryptographic} are produced from \emph{encryption}.
When using ENN we obtain completely different results where the input word is returned more than 98\% of the time, even for the lowest $\varepsilon$ value considered.

\subsubsection{Example 2: Entire Vocabulary}
\begin{figure}
	\centering
	\includegraphics[clip, width=0.6\columnwidth]{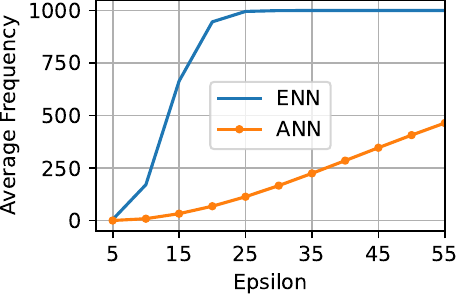}
	\caption{Average empirical frequency of the input word being output by the mechanism (out of 1000).}
	\label{fig:replication_expe2}
\end{figure}
Extending from this three-words example, in another experiment the authors consider the 319,774 common words between the two embedding models GloVe and FastText~\cite{bojanowski2017Enriching}. Each word is sanitized 1,000 times and \cite[Figure 3(b)]{feyisetan2020Privacy} plots the average for all words of the empirical frequency of the input word being returned, depending on $\varepsilon$. 
We reproduced this experiment in \cref{fig:replication_expe2} for both nearest neighbor variations. With ANN our results are similar to those in Feyisetan et al.~\cite{feyisetan2020Privacy} where for $\varepsilon$ values of 25, 35 and 50 we obtain an average frequency of $\approx 110$, $\approx 220$ and $\approx 400$ respectively.
With ENN however, the trend is totally different with the input word being returned more than 99\% of the time on average for $\varepsilon>=25$.

We believe this issue important to be raised since the $\dx$-privacy mechanism for text described in \cite{feyisetan2020Privacy, qu2021Natural} is not defined as relying on an approximation of the nearest neighbor and tends to indicate the use of an exact neighbor. Feyisetan et al.~\cite{feyisetan2020Privacy} might have chosen ANN in their implementation for the sake of performances and forgot to acknowledge it in the paper. However, we have shown this choice to heavily impact the frequency of the output being the same as the input which affects the privacy guarantee. 
Following this observation, we use a modified version of the algorithm provided by Asghar et al.~\cite{asghar2024dxprivacy} who formally studied the behavior of the mechanism with Exact Nearest Neighbor and proposed a fix consisting of lines \textit{6} and \textit{7} of \cref{alg:dx}.

\subsubsection{Sanitization Parameters}
The sanitization mechanism relies on an embedding model $\embmod$ to convert text into $n$-dimensional vectors. Word embedding models such as GloVe~\cite{pennington2014GloVe} and word2vec~\cite{word2vec} are commonly used for this task~\cite{fernandes2019Generalised, yue2021Differential, feyisetan2020Privacy, asghar2024dxprivacy} yet they have a significant limitation: words not present in their vocabulary cannot be sanitized. Simply removing such words is not an option either, as the privacy guarantees of the sanitization mechanism apply only to strings of the same length~\cite{feyisetan2020Privacy, mattern2022limits}.
This issue is particularly problematic because out-of-vocabulary words—those not encountered during the embedding model’s training—are often rare names of people or places, which are precisely the types of sensitive information that require protection.
For this reason, we opted for token embedding models from two of our language models, namely bart-large-cnn (50,264 tokens, 1024 dimensions) and Llama-3-8B-Instruct (128,256 tokens, 4096 dimensions). In our approach, text is first tokenized, and the corresponding vector for each token is used for sanitization.
Note that because the embedding models are different, the $\varepsilon$ values used throughout \cref{sec:expe} are not directly comparable with the ones in \cref{tab:dxexpe1} and \cref{fig:replication_expe2}.

\section{Experimental Results}\label{sec:expe}
In this section, we first evaluate the performance of the language models on original prompts. Next, we show the impact of prompt sanitization on the performance of all the language models. We then evaluate the capability of our utility assessor to estimate LLM performance on sanitized prompts through a regression.

\subsection{Model Performance on Original Prompts}
\begin{table}
	\centering
	\begin{subtable}{\columnwidth}
		\centering
		\begin{tabular}{|c|c|c|c|}
			\hline
			\textbf{Model} & \textbf{Avg similarity} & \textbf{stddev} & \textbf{Q10}\\ \hline
			T5 & 0.78  & 0.10 & 0.64 \\ \hline
			Bart & 0.79 & 0.10 & 0.67 \\ \hline
			Llama3.2-1B & 0.80 & 0.12 & 0.67 \\ \hline
			Llama3-8B & 0.86 & 0.08 & 0.77 \\ \hline
			Gemini & 0.83 & 0.08 & 0.73 \\ \hline
		\end{tabular}
		\caption{Summarization Task.}
		\label{tab:ogperfSumm}
	\end{subtable}
	\begin{subtable}{\columnwidth}
		\centering
		\begin{tabular}{|c|c|c|c|}
			\hline
			\textbf{Model} & \textbf{Avg similarity} & \textbf{stddev} & \textbf{Q10}\\ \hline
			T5 & 0.60  & 0.25 & 0.24 \\ \hline
			Opus-MT & 0.89  & 0.07 & 0.82 \\ \hline
			Llama3.2-1B & 0.84 & 0.12 & 0.71 \\ \hline
			Llama3-8B & 0.92 & 0.05 & 0.87 \\ \hline
			Gemini & 0.92 & 0.07 & 0.88 \\ \hline
		\end{tabular}
		\caption{Translation Task.}
		\label{tab:ogperfTrans}
	\end{subtable}
	\caption{Performance as the similarity between the original prompt and a result generated from this prompt.}
	\label{tab:ogperf}
\end{table}

Reported in \cref{tab:ogperf} is the average, standard deviation and 10th percentile of the similarity between the original prompt and a result generated from this prompt by the language models, either for summarization in \cref{tab:ogperfSumm} or translation in \cref{tab:ogperfTrans}.
As expected, the LLMs produce better results than the SLMs: for summarization Llama3-8B leads to a similarity over $0.77$ for 90\% of the samples while the best-performing SLM only produces $0.67$. For translation, Gemini has a Q10 of $0.88$ when the best SLM only yields $0.82$.

This experiment shows that while some SLMs are specialized in their respective task, the larger and general-purpose Llama-3 and Gemini still yield greater performance, justifying the use of bigger models.

\subsection{Sanitization Impact on Performance}\label{subsec:sanitizationperf}
\begin{figure*}
	\centering
	\begin{subfigure}{.5\columnwidth}
		\includegraphics[width=\columnwidth]{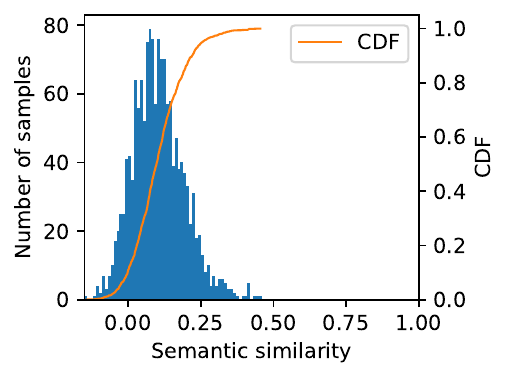}
		\caption{$\varepsilon=1$}
		\label{fig:LLAMA-OGtextVSnoisygenSummary1}
	\end{subfigure}%
	\begin{subfigure}{.5\columnwidth}
		\includegraphics[width=\columnwidth]{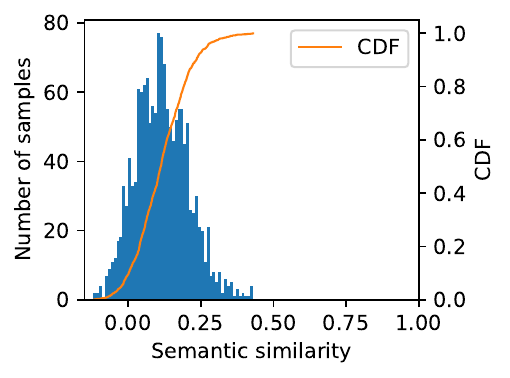}
		\caption{$\varepsilon=200$}
		\label{fig:LLAMA-OGtextVSnoisygenSummary200}
	\end{subfigure}
	\begin{subfigure}{.5\columnwidth}
		\includegraphics[width=\columnwidth]{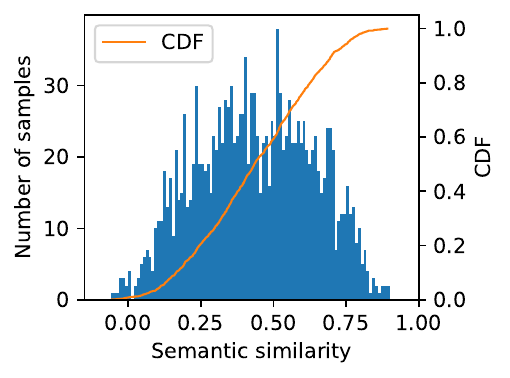}
		\caption{$\varepsilon=500$}
		\label{fig:LLAMA-OGtextVSnoisygenSummary500}
	\end{subfigure}%
	\begin{subfigure}{.5\columnwidth}
		\includegraphics[width=\columnwidth]{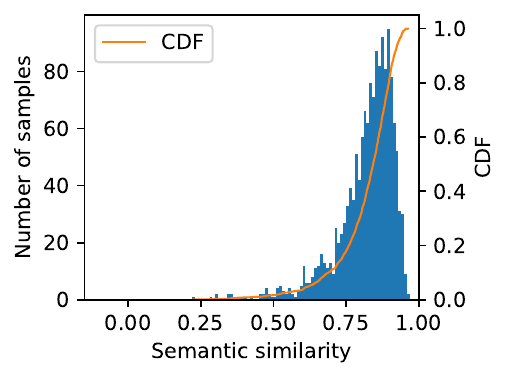}
		\caption{$\varepsilon=1000$}
		\label{fig:LLAMA-OGtextVSnoisygenSummary1000}
	\end{subfigure}
	\caption{Llama-3-8B summarization performance on sanitized prompts measured as the similarity between the original prompt and a summary generated from a sanitized version of the prompt (Sanitization with Llama's token embedding model).}
	\label{fig:LLAMA-OGtextVSnoisygenSummary}
\end{figure*}
As explained before, modifications performed by sanitizing prompts will impact their utility. In \cref{fig:LLAMA-OGtextVSnoisygenSummary}, we show the similarity between a prompt and a summary generated by Llama-3 from a sanitized version of said prompt, for different values of $\varepsilon$. The sanitization employs the embedding model of Llama3. Recall from \cref{sec:backDX} that the epsilon values presented here are not directly comparable with those usually found in ordinary DP. 

First, we notice no significant performance improvement between $\varepsilon=1$ (\cref{fig:LLAMA-OGtextVSnoisygenSummary1}) and $\varepsilon=200$ (\cref{fig:LLAMA-OGtextVSnoisygenSummary200}). This is due to the fact that the noise is relatively high in both cases and the prompts are greatly altered. 
Also, \cref{fig:LLAMA-OGtextVSnoisygenSummary500} showcases the very large range of utility one can expect for a given $\varepsilon$ value. Indeed, with $\varepsilon=500$ 20\% of the prompts lead to a similarity under 0.25 while another 20\% lead to a similarity greater than 0.62.
Finally, with $\varepsilon=1000$ the performance gets closer to the one on the original prompts with an average similarity of $0.82$, standard deviation $0.10$ and 10-th quantile of $0.69$.
Note that similar trends were obtained with the other language models and the second embedding model, and thus are not included here.
These results highlight the need for a method to estimate the utility of a given sanitized prompt before committing to use the LLM, both because the values of $\varepsilon$ are atypical but also because they affect the language modeling task to various degrees depending on the exact prompt.

\begin{figure*}
	\centering
	\begin{subfigure}{.5\linewidth}
		\includegraphics[width=\columnwidth]{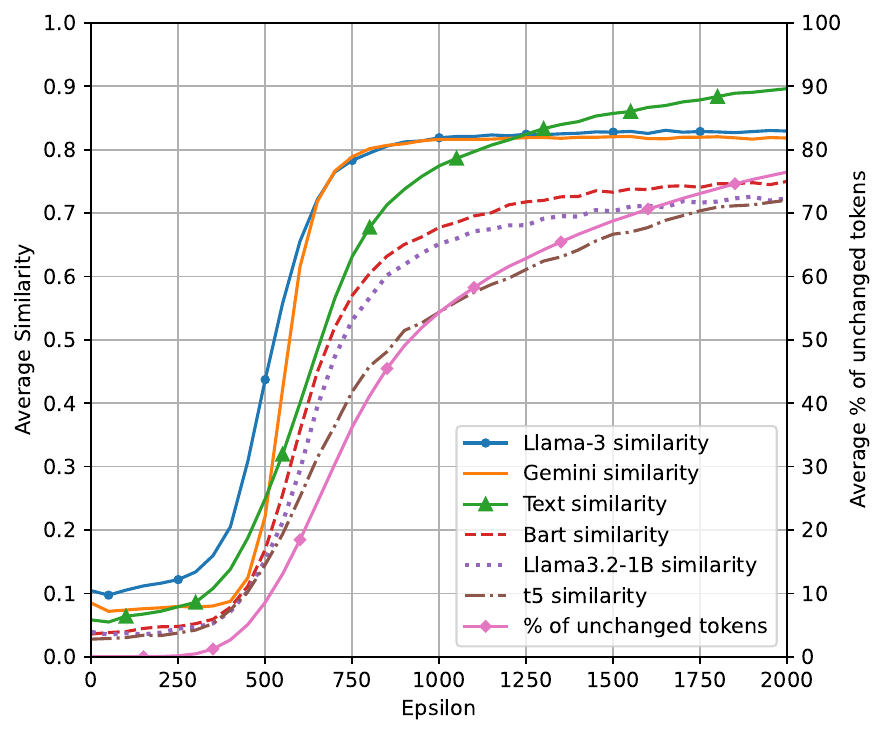}
		\caption{Sanitization with Llama's token embedding model}
		\label{fig:AllModelsPerfs-LlamaEmbMod}
	\end{subfigure}%
	\begin{subfigure}{.5\linewidth}
		\includegraphics[width=\columnwidth]{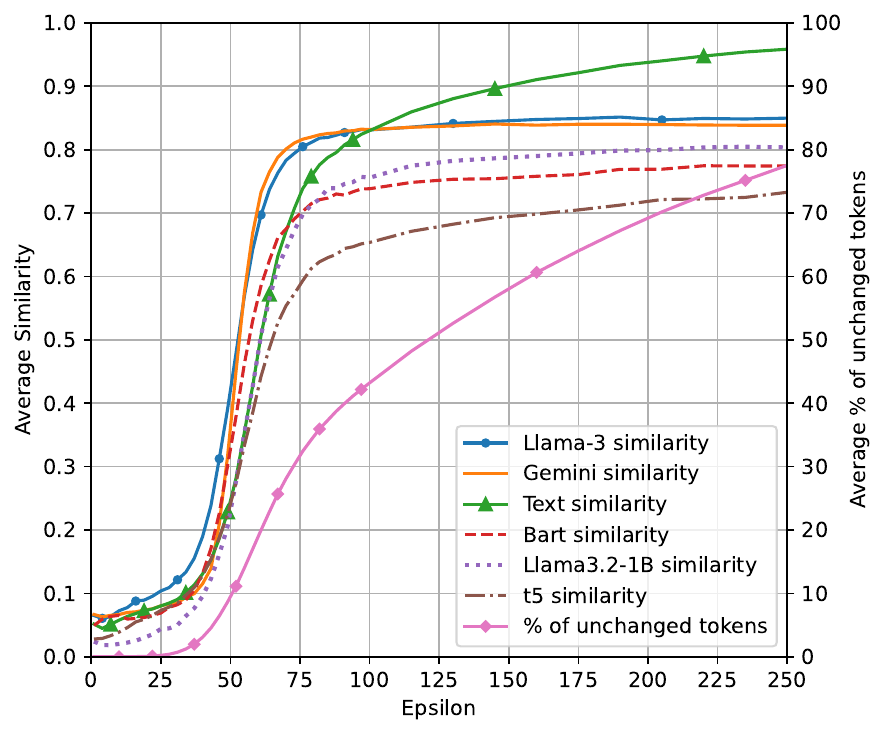}
		\caption{Sanitization with Bart's token embedding model}
		\label{fig:AllModelsPerfs-BartEmbMod}
	\end{subfigure}
	\caption{Models summarization performance on sanitized prompts, compared with the percentage of tokens changed and the similarity between the original prompt and a sanitized version of the prompt.}
	\label{fig:AllModelsPerfs}
\end{figure*}
We now turn our attention to the multiple language models under evaluation, with
\cref{fig:AllModelsPerfs} providing an overview of the summarization performances of each language model considered for this task as a function of $\varepsilon$. In \cref{fig:AllModelsPerfs-LlamaEmbMod} the sanitization involves Llama's embedding model while \cref{fig:AllModelsPerfs-BartEmbMod} uses Bart's embedding model. In these charts "Text Similarity" is the similarity between a prompt and a sanitized version of this prompt, averaged for all prompt at each $\varepsilon$ value. "\% of unchanged tokens" represents the average percentage of tokens per prompt that have not been modified. 
Finally, the performance of each language model is represented as the average similarity between the original prompt and a summary generated by the model from a sanitized version of the prompt.

In \cref{fig:AllModelsPerfs-LlamaEmbMod}, from $\varepsilon=1$ to $250$ all models have equally poor performance when generating summaries from highly sanitized prompts, as reflected in the summarization similarities ranging from $0.04$ (Llama3.2-1B) to $0.12$ (Llama3). In this interval the noise is too high to get any utility: $99.9$\% of tokens are modified and the sanitized prompts are extremely dissimilar with their original version (similarity=$0.06$). 
Next, from $\varepsilon=300$ to $1250$, the texts' similarity increases to $0.82$, which greatly assists the language models in summarization. Llama-3 and Gemini display the best performance progression, reaching a similarity above $0.7$ at $\varepsilon=650$ while Bart, Llama3.2-1B and T5 attain this value at $\varepsilon=1150$, $\varepsilon=1450$ and $\varepsilon=1750$ respectively. Beyond $\varepsilon=1000$ the performance of the two LLMs plateau.

In \cref{fig:AllModelsPerfs-BartEmbMod}, while the range of $\varepsilon$ values is different because of the embedding model, similar trends are represented: Highly sanitized prompts (i.e. $\varepsilon\in[1,25]$) produce mediocre performance across all models, then LLMs have sharper performance increase than SLMs until stabilizing at $\varepsilon=75$.

Finally, we note that language models are generally better at summarizing when Bart's embedding model is involved in the sanitization instead of Llama's. For example, when 20\% of the tokens are unchanged by the sanitization, the average similarity for all models is $0.59$ for Bart's embedding model and $0.51$ for Llama's. While the difference in dimensionality and number of tokens may have an impact, we leave experiments with other embedding models to study this trend for future works.
For the translation task, similar results were obtained and figures are omitted here for brevity.

While the LLMs clearly remain better at handling sanitized prompts, the performance trends between all models have similar characteristics. This provides us with a solid basis for our ultimate objective: predicting the performance of an LLM based on the performance of an SLM.

\subsection{Predicting LLM Performance via Regression}\label{subsec:predi}
When considering the similar performance trends between our SLMs and LLMs when processing sanitized prompts, we devise the following experiment to address our problem statement.
We consider the following set of features derived from our utility assessor architecture:

\begin{enumerate}[label=\Alph*)]
	\item A prescribed value of $\varepsilon$,
	\item The semantic similarity between the original prompt and the sanitized prompt,
	\item The semantic similarity between the original prompt and a summary/translation generated from said prompt by an SLM,
	\item The semantic similarity between the original prompt and a summary/translation generated from a sanitized version of the prompt by an SLM,
	\item (Target feature) The semantic similarity between the original prompt and a summary/translation generated from a sanitized version of the prompt by an LLM.
\end{enumerate}

Given features A to D, we perform a regression to predict feature E. We present the results in \cref{tab:predictionSumma} for summarization and in \cref{tab:predictionTranslation} for translation. Each row involves its own combination of embedding model for the sanitization, LLM, and SLM. As our baseline, we consider a regression solely based on feature A i.e., leveraging the information available without our utility assessor.

This regression is performed on 1500 random samples of our dataset. For summarization, each prompt is sanitized with $\varepsilon$ varying from 1 to 1000 with a step of 50 for Llama's embedding model, and from 1 to 100 with a step of 3 for Bart's embedding model. We focus on this particular ranges of epsilons because, as seen in \cref{fig:AllModelsPerfs}, $\varepsilon$ does impact LLMs' performance in this range while it does not impact it significantly for higher values. In our context, we believe there is no reason for a user to opt for higher epsilon values (i.e., lesser privacy) if there is no gain in performance.
Following the same reasoning, for translation the $\varepsilon$ values range from 1 to 2000 with a step of 50 for Llama's embedding model, and from 1 to 250 with a step of 3 for Bart's embedding model.

We used a Histogram-based Gradient Boosting Regression Tree from the scikit-learn library~\cite{scikit-learn} with a Train/Test split of $80\%$/$20\%$. Other regressors were tested such as Decision Tree or Support Vector. We selected the best-performing one.

The Llama 3.2-1B SLM has two variants. In its first variant, it performs the summarization on the output of the sanitization procedure, exactly as the other SLMs do. In the second variant (Corrected Prompts = Yes), we benefit from the fact that this SLM is general-purpose and we first leverage it to correct the prompts for coherence and grammar. Then, both the SLM and the LLM perform the task on the corrected prompts.

The performance of the regressor is measured in terms of average precision with the coefficient of determination ($R^2$) and the Root Mean Square Error (RMSE). Furthermore, we include a measure specific to our context: "Wasted \$" denotes the percentage of prompts where resources are wasted because the prediction was too optimistic. This is counted with the number of prompts where the regressor predicted a similarity of more than $0.1$ higher than the actual target value. Also, "Failed Prediction" denotes the percentage of prompts where the prediction is overly optimistic or overly pessimistic, counted as the number of prompts where the regressor predicted a similarity having an absolute difference greater than $0.1$ with the target value.

The general trend shows the benefit of our middleware where it always improves the regression performance on both tasks over the baseline regardless of the SLM. In general, the more parameters an SLM has the better the prediction, with Llama3.2-1B as our top contestant. The best improvement reduces failed predictions by 53\% compared to the baseline, 20\% of which were wasting computing and monetary resources (See \cref{tab:predictionTranslation} with Llama's embedding model, Gemini as the LLM and Llama3.2-1B SLM on corrected prompts).
On average for both tasks, using Llama3.2-1B on corrected prompts reduces failed prediction by 21\% and saves 10\% of wasted resources.
Even with an extremely small model like T5 (60M parameters), failed prediction are reduced by 14\% on average. 

\renewcommand{\arraystretch}{1.1} 
\begin{table*}
	\centering
	\begin{subtable}{\textwidth}
		\centering
		\begin{tabular}{|c|c|c|c|c|c|c|c|c|}
			\hline
			\makecell{\textbf{Embedding}\\\textbf{Model}} & \textbf{LLM} & \textbf{SLM} & \makecell{\textbf{Corrected}\\\textbf{Prompts}} & \textbf{Features} &\textbf{R²} & \textbf{RMSE} & \textbf{Wasted \$} & \makecell{\textbf{Failed}\\ \textbf{Prediction}}\\ 
			\hline
			Llama  & Llama 3-8B & \emph{None} & No & A & 0.85 & 0.13 & 19\% &  36\% \\
			& & T5 & No & ABCD & \textbf{0.89} & 0.11 & 16\% &  30\% \\ 
			&   & Bart & No & ABCD & \textbf{0.89} & 0.11 & 14\% & 28\% \\ 
			&   & Llama 3.2-1B & No & ABCD & \textbf{0.89} & 0.11 & 15\% & 28\%\\ 
			&   & Llama 3.2-1B & Yes & ABCD & \textbf{0.89} & \textbf{0.08} & \textbf{7\%} & \textbf{16\%} \\ 
			\cline{2-9}
			& Gemini & \emph{None} & No & A & 0.87 & 0.13 & 15\% & 30\% \\
			&   & T5 & No & ABCD & \textbf{0.91} & 0.11 & 11\% & 23\%\\ 
			&   & Bart & No & ABCD & \textbf{0.91} & 0.10 & 11\% & 22\%\\ 
			&   & Llama 3.2-1B & No & ABCD & \textbf{0.91} & 0.11 & 10\% & 20\%\\
			&   & Llama 3.2-1B & Yes & ABCD & 0.89 & \textbf{0.08} & \textbf{6\%} & \textbf{14\%} \\ 
			\hline
			\hline
			Bart  & Llama 3-8B  & \emph{None} & No & A & 0.88 & 0.12 & 17\% &  33\%\\
			& & T5 & No & ABCD & 0.90 & 0.11 & 14\% &  27\% \\ 
			&  & Bart & No & ABCD & \textbf{0.91} & \textbf{0.10} & 14\% &  26\%  \\ 
			&  & Llama 3.2-1B & No & ABCD & \textbf{0.91} & \textbf{0.10} & 13\% &  25\%  \\ 
			&  & Llama 3.2-1B & Yes & ABCD & \textbf{0.91} & \textbf{0.10} & \textbf{11\%} & \textbf{22\%} \\ 
			\cline{2-9}
			& Gemini & \emph{None} & No & A & 0.90 & 0.11 & 14\% & 27\%\\
			& & T5 & No & ABCD & \textbf{0.93} & 0.10 & 10\% & 20\%\\ 
			&  & Bart & No & ABCD & \textbf{0.93} & \textbf{0.09} & \textbf{9\%} & \textbf{18\%}  \\ 
			&  & Llama 3.2-1B & No & ABCD & \textbf{0.93} & \textbf{0.09} & \textbf{9\%} & \textbf{18\%}  \\
			&  & Llama 3.2-1B & Yes & ABCD & 0.92 & 0.10 & 10\% & 21\%  \\ 
			\hline
		\end{tabular}
		\caption{Summarization task.}
		\label{tab:predictionSumma}
	\end{subtable}
	
	\begin{subtable}{\textwidth}
		\centering
		\begin{tabular}{|c|c|c|c|c|c|c|c|c|}
			\hline
			\makecell{\textbf{Embedding}\\\textbf{Model}} & \textbf{LLM} & \textbf{SLM} & \makecell{\textbf{Corrected}\\\textbf{Prompts}} & \textbf{Features} &\textbf{R²} & \textbf{RMSE} & \textbf{Wasted \$} & \makecell{\textbf{Failed}\\ \textbf{Prediction}}\\ 
			\hline
			Llama  & Llama 3-8B & \emph{None} & No & A & 0.76 & 0.16 & 23\% & 56\% \\
			&   & T5 & No & ABCD & 0.87 & 0.12 & 13\% & 27\% \\
			&   & Opus-MT & No & ABCD & 0.88 & 0.12 & 12\% & 24\% \\
			&   & Llama 3.2-1B & No & ABCD & 0.88 & 0.12 & 12\% & 24\%\\ 
			&   & Llama 3.2-1B & Yes & ABCD & \textbf{0.95} & \textbf{0.07} & \textbf{8\%} & \textbf{15\%} \\ 
			\cline{2-9}
			& Gemini & \emph{None} & No & A & 0.60 & 0.20 & 25\% & 61\% \\
			&   & T5 & No & ABCD & 0.80 & 0.14 & 14\% & 25\% \\
			&   & Opus-MT & No & ABCD & 0.80 & 0.14 & 14\% & 25\% \\
			&   & Llama 3.2-1B & No & ABCD & 0.80 & 0.14 & 13\% & 24\%\\
			&   & Llama 3.2-1B & Yes & ABCD & \textbf{0.96} & \textbf{0.06} & \textbf{5\%} & \textbf{8\%} \\ 
			\hline
			\hline
			Bart  & Llama 3-8B  & \emph{None} & No & A & 0.92 & 0.10 & 13\% & 24\%\\ 
			&   & T5 & No & ABCD & \textbf{0.95} & 0.08 & 8\% & 15\% \\
			&   & Opus-MT & No & ABCD & \textbf{0.95} & 0.08 & 8\% & 14\% \\
			&  & Llama 3.2-1B & No & ABCD & \textbf{0.95} & \textbf{0.07} & 8\% & 14\%  \\ 
			&  & Llama 3.2-1B & Yes & ABCD & \textbf{0.95} & \textbf{0.07} & \textbf{6\%} & \textbf{12\%} \\ 
			\cline{2-9}
			& Gemini & \emph{None} & No & A & 0.93 & 0.09 & 13\% & 23\%\\           
			&   & T5 & No & ABCD & \textbf{0.97} & \textbf{0.06} & 5\% & 9\% \\
			&   & Opus-MT & No & ABCD & \textbf{0.97} & \textbf{0.06} & 5\% & 8\% \\
			&  & Llama 3.2-1B & No & ABCD & \textbf{0.97} & \textbf{0.06} & 5\% & 8\%  \\
			&  & Llama 3.2-1B & Yes & ABCD & \textbf{0.97} & \textbf{0.06} & \textbf{4\%} & \textbf{7\%}  \\ 
			\hline
		\end{tabular}
		\caption{Machine Translation task.}
		\label{tab:predictionTranslation}
	\end{subtable}
	\caption{Regression Performances. Rows with feature A only represent the baseline i.e., a regression without our architecture. "Wasted \$" denotes the percentage of prompts where the prediction is overly optimistic (i.e. $\text{target}<\text{prediction}-0.1$) and resources are wasted. "Failed Prediction" includes the previous column and denotes the percentage of prompts where the prediction is overly optimistic or overly pessimistic (i.e. $abs(\text{target}-\text{prediction})>0.1$).}
	\label{tab:prediction}
\end{table*}
\renewcommand{\arraystretch}{1.0}

\section{Related Works}\label{sec:sota}
In this section, we first present related works on text sanitization methods other than the ones based on DP which were described in \cref{sec:backDX}. 
Then, we give an overview of works related to the privacy of LLM prompts.

\subsection{Text Sanitization}
Extensive labor is required for humans to manually redact textual documents \cite{gordon2013mra}. As a consequence automatic methods from the Natural Language Processing (NLP) domain have been proposed, resulting in solutions that either sanitize specific sensitive components of the text detected beforehand, or the entire text.

\subsubsection{PII Detection}
With the advent of Machine Learning for NLP, popular frameworks/libraries such as spaCy~\cite{Honnibal_spaCy_Industrial-strength_Natural_2020} or Presidio~\cite{microsoftPresidio} have emerged with functionalities to perform Named Entity Recognition (NER) for the detection of Personally Identifiable Information (PII) in textual data \cite{chen2023Hide, chong2024Casper}.  
For example, entities can be related to persons (names, DOB, address etc.), locations or organizations. 
\cite{hassan2023Utilitypreserving} proposes an alternative to NER by training a word embedding model on sensitive documents to capture the relation between words and entities that we want to protect. In addition to NER, \cite{papadopoulou2023Neural} exploits ``gazetteers'' which are collections of terms under specific groupings (e.g. job titles, educational background, part of their physical appearance etc.), created from knowledge graphs such as WikiData~\cite{wikidata}, which they argue are not captured by NER but are nonetheless quasi-identifying.

After PII has been detected, different solutions have been proposed to sanitize PII, either through simple methods of complete removal~\cite{papadopoulou2023Neural} or placeholders~\cite{chen2023Hide}. For more advanced methods, the PII can be sanitized via Differential Privacy (See \cref{sec:backDX}), Generalization or Rewriting (See below).

\subsubsection{Sanitization via Generalization}
Generalization sanitizes certain types of PII by using a more general and less discriminating term~\cite{hassan2023Utilitypreserving, olstadGenerationSelectionReplacement2023}. PII in a text are searched against a knowledge base such as WikiData~\cite{wikidata}, and generalization exploits metadata of the base in order to sanitize the PII. For example, the name of a city is replaced by ``a city of X'' where X is a name of a country, or a specific company name is replaced by ``a telecommunication company''. The privacy trade-off is configured by generalizing to a lower or higher degree when possible.
This solution is limited in that it applies to quasi-identifiers but not to direct identifiers. Also, this approach is heavily dependent on the exhaustiveness of the knowledge base, as such, blind spots may arise for specific, field-related terms not common in public knowledge.

\subsubsection{Private Rewriting}
Language models can also be leveraged to automatically rewrite a text with added privacy properties. \cite{weggenmann2022DPVAE} relies on variational autoencoders, a specific type of autoencoder where the output of the encoder is a probability distribution over the embedding space instead of a particular point. In a nutshell, an entire text is given to the encoder which produces several embedding vectors for the entire text, associated with a probability distribution.
One embedding vector is then sampled from this distribution and provided to the decoder to reconstruct the input in place of the original. \cite{weggenmann2022DPVAE} modifies the encoder to produce a specific probability distribution providing differentially-private properties.
\cite{igamberdiev2023DPBART} on the other hand take a standard encoder-decoder model (BART~\cite{lewis2019BART}) but add noise to the encoder's output, before it is passed to the decoder to generate sanitized text.
Both of these contributions apply a predetermined amount of differentially-private noise without regard for utility implications, and would benefit from a means to assess the utility of a given rewritten text, which we study in this paper.

\subsection{Privacy for LLM Prompts}
As the sanitization of prompts induces a direct impact on utility, one cannot have both high utility and high privacy on any given prompt \cite{zhang2024No}. In this section we mention related works for prompts privacy and how they navigate the privacy-utility trade-off.
\subsubsection{Hide \& Seek}
\cite{chen2023Hide} propose a system to anonymize PII in prompts sent to an LLM and consequently de-anonymize the results. The authors propose two anonymization schemes: generative and label-based. The generative scheme features a training phase where a corpus of prompts is anonymized by an online LLM requested to replace particular private entities with ``random words''~\cite{chen2023Hide}. Then, a smaller language model is tasked to anonymize in the same manner by providing it both the original prompts and the corresponding LLM anonymized prompt.
The label-based scheme uses pre-trained NER models to detect PII and replace it with placeholders dependent on the PII's category, e.g. \textless DATE\textgreater\ for a date.
By replacing entities with random words or placeholders, both schemes do not preserve semantics of the PII. Also for the generative scheme it is unclear to what extent the smaller language model will be able to anonymize entities not seen during training. The label-based scheme is dependent on the NER models to correctly detect entities for sanitization. Finally, we are not aware of a means to navigate the privacy-utility trade-off similar to the $\varepsilon$ value in DP. In our work, we use a text sanitization mechanism where the semantic is preserved through related tokens and the privacy-utility trade-off is navigable through our utility assessor architecture.

\subsubsection{ProSan}
The authors of \cite{shen2024fire} propose a solution to sanitize the prompts of LLMs. Each word of a prompt is assigned an importance score and a privacy risk. By using a second LLM (separate to the untrusted LLM that is to be queried), the importance score is computed through the gradient of the loss function to measure how much the output will be changed if the given word is modified. The privacy risk also leverages an LLM, by quantifying how improbable a given word was according to the language modeling head.
The words to be sanitized are the first $\gamma$ least important words, $\gamma$ being adjusted depending on the overall privacy leakage risk of the prompt (quantified by its entropy).
Replacement words are chosen from a list generated by a Mask Language Model. 
This framework modifies the words that have a low importance score and high privacy sensitivity. However, we argue that by specifically targeting low importance words the semantic of the prompt is indeed preserved, but the arguably more significant case of words having both a high importance score and a high privacy sensitivity is not considered. In our work, all words of a text are sanitized independently and thus the privacy guarantees of the sanitization do not depend on the correct detection of sensitive terms.

\subsubsection{InferDPT}
\cite{tong2024InferDPT} is similar to our work where prompts are sanitized using a differentially-private mechanism parameterized by $\varepsilon$. A Small Language model is locally hosted and the user wants to query an online and untrusted LLM to perform a text continuation task on a sensitive document. The LLM performs the task on a sanitized version of the document and the SLM leverages both the original document and the output of the LLM to produce the final result. First, their system always queries the LLM even when the document has lost all utility due to strong sanitization, a practice we argue is wasting monetary and computing resources which we specifically try to avoid in this paper. Moreover, some of the experiments results are incoherent where their system is measured as having better performance than GPT4 alone on the original document and since the codebase is not public, we cannot reproduce their results.

\subsubsection{With LLM's Help}
Other research works benefit from the collaboration of the online LLM itself to perform private inference. 
For example, \cite{mai2023SplitandDenoise} explores a context where the Embedding layer of a language model is available at user-side, thus allowing the addition of DP noise to embeddings without mapping them back to tokens. 
In \cite{qu2021Natural} the authors show that pre-training a language model with sanitized data assists the model achieve better performance during inference.
In this paper we focus on contexts where the LLM to be queried is untrusted and black-box, and sanitization have to be performed in a text to text manner to maintain compatibility with the pipeline of current, general-purpose online LLMs.

\section{Discussions and Limitations}\label{sec:limitations}
\subsection{Extending to Other Mechanisms and Tasks}
In this paper, we experimented with a token-level sanitization mechanism based on DP presented in \cref{sec:backDX}. Yet, other sanitization mechanisms such as LLM-assisted Private Rewriting~\cite{igamberdiev2023DPBART} will also trade utility for privacy and as such will benefit from our architecture to better navigate this tradeoff. Our architecture can be used as long as the sanitization mechanism produces text as its output.

Generalization to other language model tasks depends on the availability of a metric to measure the quality of a result without a reference (or ground-truth). Indeed, the Quality Estimator of \cref{fig:architecture} must evaluate $p_\varepsilon$, $r^{SLM}$ and $r_\varepsilon^{SLM}$ without having such a reference. As shown in \cref{sec:expe}, cosine similarity with the original prompt is an example of such metric which can be applied on a summarization and a machine translation task. Other language model tasks can be envisioned:
Text Classification can be evaluated based on the confidence score of the SLM, or Open-ended Text Continuation with coherence and/or diversity. However, Named Entity Recognition and Question Answering are examples of tasks where, to the best of our knowledge, reference-free quality estimation appears difficult.

\subsection{Middleware Impact on DP Guarantees}
In this paper, we assume that a prompt is processed only once through the middleware, and any two prompts submitted by the user are substantially different. In case of similar prompts, the middleware can reuse the previously saved sanitized version. This assumption is essential to ensure that the LLM provider does not see several sanitized versions of the same prompt, which could lead to an averaging attack similar to the one sketched in \cref{sec:sanit_perf_consideration}.

Having said that, to assess the quality of the sanitized prompt, the middleware uses the original, sensitive prompt. Since this assessment is not differentially private, technically speaking, the decision to send the prompt to the LLM is not differentially private as a whole. To understand this, consider an extreme example where the user's prompt consists of only one token. The middleware may decide that the only ``sanitized'' version that satisfies the quality criterion is the original token itself. In this case, privacy is blatantly breached, as the LLM, knowing the decision criteria of the middleware, can guess that the received token is highly likely to be the original token. 

While this attack is possible in theory, in practice user prompts are expected to be sufficiently long, in which case it is difficult for the remote LLM to ascertain which tokens have been changed and which ones remain unaltered. Consider the case where quality is measured by cosine similarity and assume that the LLM knows the parameters of the middleware, including $\epsilon$ values and threshold for similarity. Upon receiving the sanitized prompt, the LLM only learns that the embedding vector of the sanitized prompt (i.e., the vector resulting from the average of the embeddings of its tokens) has a similarity with the embedding vector of the original prompt above the threshold. It does not learn the exact similarity value, and more importantly, the degree to which each token has participated in that similarity, given that each token is sanitized separately.

\section{Conclusion}\label{sec:concl}
To conclude, this paper studies the utility impact of text sanitization when applied on Large Language Model (LLM) prompts. 
Our goal is to conserve resources, both monetary for users and computational for the LLM provider, by predicting when sanitized prompts have lost their effectiveness for generating meaningful LLM responses. This is particularly relevant in a professional context where a company benefits from employees using LLMs to increase productivity, but at the same time wants to avoid leakage of trade secrets and company data in the prompts. To do so they will resort to automatic sanitization methods.

We focus on a Differential Privacy (DP) framework applied on text called $\dx$-privacy. We show that it is difficult to anticipate the performance of an LLM on a sanitized prompt, and thus to prevent resource waste. We propose a middleware architecture including a locally trusted Small Language Model (SLM) to inform users about the utility/privacy trade-off before engaging with the LLM.

We explore the potential of this architecture by conducting summarization and translation tasks with a publicly available dataset, two LLMs, and five SLMs of different capabilities.
Our experiments are clear on the heavy and unintuitive impact sanitization can have on the utility of prompts. 
Nonetheless, we show that the performance trends between SLMs and LLMs are similar. By leveraging numerical features available via our middleware, we train a regression model and confirm the benefit of our architecture where up to 20\% of the prompts previously wasting resources are now saved. 
In a company setting this is 20\% of all API calls saved, while precious computing resources are also saved on the LLM's provider side.

During our study, we also shed light on an issue with $\dx$-privacy applied on text: an implementation strictly following the definition of one of its most-cited paper~\cite{feyisetan2020Privacy} has a severely degraded privacy protection.

We acknowledge the limitations of our middleware, mainly its support for only those language model tasks whose quality can be evaluated without a reference, a criterion we argue already includes important tasks such as summarization and translation. Also, since prompts of poor quality are not sent in order to save resources, the LLM provider knows that a received prompt passed a certain quality threshold, which weakens the DP guarantees of the sanitization. We argue that in practice this information is of little use to the provider to break the privacy of the tokens.


\bibliographystyle{ACM-Reference-Format}
\bibliography{references}


\end{document}